\magnification=\magstep1
\parindent=0pt
\parskip=20pt
\font\tensb=msbm10
\def\bbC{\hbox{\tensb C}}
\def\Cstar{\bbC^*}
\def\hook{\hbox{\vrule height0pt width4pt depth0.5pt
\vrule height7pt width0.5pt depth0.5pt \vrule height0pt width2pt
depth0pt} }
\def\del#1{\nabla_{X_{#1}}}
\def\Tr{\hbox{Tr }}
\def\nablahat{\widehat{\nabla}}
\def\dhat{\hat{d}}
\def\codhat{\hat{d^*}}

\def\Dhat{\hat D}
\def\gradhat{\widehat{\rm grad}}



\pageno=-1
\centerline{\bf Debye Potentials for Maxwell and Dirac Fields }
\vskip 0.2true cm
\centerline{\bf from a}
\vskip 0.2true cm
\centerline{\bf Generalisation	of the Killing-Yano Equation.}
\vskip 0.3true cm
\centerline{\bf }
\vskip 1true cm
\centerline{ {\bf I. M. Benn}\footnote{${}^1$}
{email:mmimb@cc.newcastle.edu.au}}
\centerline{{\bf Philip Charlton}\footnote{${}^2$}
{email:jkress@maths.newcastle.edu.au}}
\centerline{{\bf Jonathan Kress}\footnote{${}^3$}
{email:philipc@maths.newcastle.edu.au}}
\vskip 0.5true cm
$$\vbox{\halign{\hfil#\cr        
Mathematics Department,\cr
Newcastle University,\cr
NSW 2308,\cr
Australia.\cr}}$$

\vskip 1true cm
\noindent{\bf Abstract.} By using conformal Killing-Yano tensors, and their
generalisations, we obtain scalar potentials for both the source-free Maxwell
and  massless Dirac equations. 
For each of these equations we construct, from conformal Killing-Yano 
tensors,  symmetry operators  that map any solution to another.

\vfill\eject
\pageno=1
\noindent{\bf I. Introduction.}

Maxwell's equations require that the Maxwell 2-forms be closed, and hence 
locally	exact. They are globally exact if we discount the existence of
magnetic monopoles, and Maxwell's equations can then be written 
as second-order equations for  a potential 1-form. 
In the absence of sources the Maxwell 2-forms are also co-closed and hence
locally co-exact. For charge-free solutions they are globally co-exact and
a potential 3-form can be introduced. A Hertz potential is a 2-form 
in terms of which  the Maxwell 2-form
can be expressed so as  to be simultaneously exact and co-exact, and 
hence
satisfy the source-free Maxwell equations. In certain cases one can
parameterise the Hertz potential in terms of a function satisfying 
a second-order differential equation. Thus Maxwell solutions can be
expressed in terms of a scalar potential: the Debye potential.
(Clearly we have paraphrased what Hertz and Debye actually did. Although
we shall not attempt to give an historical account it might be noted
that Bromwich and Whittaker also contributed to these ideas. We have
given references later.)

Some of the important solutions to Maxwell's equations in flat space are
expressed in terms of solutions to scalar equations. If the Lorenz gauge
condition is imposed on the potential 1-form then  Maxwell's equations require
that it be harmonic. In a parallel basis this requires that its components be
harmonic functions. The plane wave solutions may be obtained in this way. The
radiating multipole solutions are adapted to spherical symmetry. Here the
electric and magnetic fields are required to satisfy a vector Helmholtz
equation. Solutions to this vector equation can be expressed in terms of
solutions to a scalar Helmholtz equation. Thus the radiating multipole family
can be expressed in terms of scalar potentials. In a curved space-time one
cannot immediately generalise these solutions, as the derivatives introduce
extra connection terms. Cohen and Kegeles [1] seem to have been the first
to apply the Hertz potential formalism to Maxwell's vacuum equations in a
curved background.  They point out that the Hertz potential must be some
geometrically privileged 2-form, and that algebraically-special space-times
have such 2-forms, corresponding to the repeated principal null directions.

Sections II to V are all preliminaries to the Debye potential formalism.
A preliminary reading may begin at section VI, referring to the earlier
sections as necessary.
Section II introduces some notation, whilst Section III expresses some
results on the Petrov classification scheme in a form that will be
convenient later. In Section IV we introduce the conformal Killing-Yano
equation. This conformal generalisation of Yano's Killing-tensor equation
was introduced in [2]. We show how the equation can be written 
equivalently in terms of exterior operations. This provides an
elegant statement of the equation.
Moreover, it is natural to have the equation expressed in this way
when considering Maxwell's equations which are themselves most naturally
expressed in exterior form.
Any differential form may be regarded as a tensor on the spinor
space. We show how the conformal Killing-Yano equation for a 2-form
is equivalent to  an equation for a 2-index Killing spinor. This relationship
has not always been made clear: indeed, tensors equivalent to Killing
spinors have been called Penrose-Floyd tensors.
In Section V we consider equations for self-dual
2-forms whose eigenvectors are
shear-free. When the 2-forms have only one real eigenvector then Robinson's
theorem [3] says that they are proportional to a closed form. We 
show that equivalently the 2-forms satisfy  a `gauged' conformal
Killing-Yano equation. This restatement of Robinson's theorem proves
convenient for obtaining Debye potentials. When the 2-form has two real
shear-free eigenvectors then we recover the equations obtained by
Dietz and R\"udiger [4]. 	
We obtain the integrability conditions for these shear-free equations
that we will need later.  We do this by relating the self-dual 2-forms
to spinor fields. This is not only a convenient way of obtaining these
integrability conditions, but also enables us to consider the Debye
scheme for the Dirac equation in section VII.

In section VI we show how Debye potentials 
%
%
for Maxwell fields
%
%
are related to repeated principal null directions in algebraically-special
space-times. This had previously been done in Newman-Penrose formalism
by Cohen and Kegeles [1, 5]. Thus their Debye potential equations were
written out explicitly in an adapted basis. The advantage of our approach
is that we obtain the equations in a basis-independent way, which we
feel makes it easier to see how the various ingredients to the scheme
enter. (Of course, to solve the equations in any given space-time 
necessitates  choosing some basis adapted to the geometry.) 
%
%
Other workers have utilised different aspects of special properties of
space-times admitting Debye potentials.  Stewart [6] 
considered Petrov type $D$ vacuum space-times
and used the more specialised Geroch-Held-Penrose formalism to 
obtain simplified equations.  Torres del Castillo [7] emphasises
the special properties of totally null foliations or `null strings' in
his treatment of null Hertz potentials.  Wald [8] has pointed out the
nature of the 
relationship between Cohen and Kegeles' Debye potential equations the 
decoupled massless field components of Teukolsky [9, 10] from 
which he derives the Debye potential equations.

%
%
In the special case in which there exists a conformal 
Killing-Yano tensor the Debye potential scheme becomes both simpler and
more powerful. 
(Here our way of writing the conformal Killing-Yano equation is 
particularly  well adapted to the scheme.) It is possible to use 
Debye potentials to obtain a symmetry
operator for Maxwell's equations, mapping any source-free Maxwell solution
to another. This has been shown by Torres del Castillo [11] using
2-index Killing spinors. Now Kalnins, McLenaghan and Williams [12]
have obtained the most general second order
symmetry operator for the source-free
Maxwell system. 
Their operator contains a term constructed from a 4-index Killing spinor.
We show how such an operator is obtained from the Debye potential scheme.
We show how it is expressed in terms of 
a `generalised conformal Killing-Yano
tensor'. 

In section VII we treat the massless Dirac system analogously  to
the Maxwell system. We show the relation to previous work, and in
particular show how the Debye potential method 
generalises Penrose's `spin raising' and `spin lowering' operators,
constructed from twistors in conformally-flat space [13], 
to an algebraically-special space-time.

In the final section we summarise our results and discuss 
possible generalisations.  

\vskip 0.3true cm
\noindent{\bf II. Notation and Conventions.}
\vskip 0.3true cm
The exterior calculus of differential forms will be used extensively.
As usual $\wedge$ denotes the exterior product and $d$ the exterior
derivative. If $X$ is a vector field then $X\hook$ denotes the interior
derivative that contracts a $p$-form with $X$ to produce a $(p-1)$-form.
The interior derivative is an anti-derivation on differential forms 
(as is $d$) with  $X\hook A=A(X)$ for $A$ a 1-form.
It follows that if $\omega$ is a $p$-form then
$$e^a\wedge X_a\hook\omega = p\omega $$
where the coframe $\{e^a\}$ is dual to the frame $\{X_a\}$.

On a pseudo-Riemannian manifold the metric tensor $g$ establishes a natural
isomorphism between vector fields and differential 1-forms. The 1-form
$X^\flat$ is defined such that $X^\flat(Y)=g(X,Y)$ for all vector fields
$Y$. Thus $X^\flat$ has components obtained by lowering those of $X$ with the
metric tensor. The inverse of $\flat$ is $\sharp$, resulting in the vector
$\omega^\sharp$ having components obtained by raising those of the 1-form
$\omega$.

The metric tensor also gives a natural isomorphism between the space
of $p$-forms and the $(n-p)$-forms, where $n$ is the dimension of the
manifold. This isomorphism is the Hodge duality map $*$. 
On a decomposable $p$-form we have
$$*(X_1{}^\flat\wedge X_2{}^\flat\wedge\ldots\wedge X_p{}^\flat	)
= X_p\hook X_{p-1}\hook\ldots X_1\hook *1 \eqno(1)$$
where  $*1$ is the orienting volume $n$-form.
{\sl We shall only consider four-dimensional \break
Lorentzian manifolds.} The metric
tensor $g$ will be taken to be positive-definite on  space-like vectors.
With these conventions we have
$$**\omega=\cases{\omega& if $\omega$ is of odd degree;\cr
-\omega& if $\omega$ is even.}\eqno(2)$$
The Hodge dual may be combined with the exterior derivative to form the
co-derivative $d^*$ that acts on a form to lower the degree by one. For 
the case of four dimensions and Lorentzian signature 
$$d^* =  *d*\,.\eqno(3) $$
The exterior derivative and the co-derivative can be expressed in terms of
the Lorentzian connection $\nabla$ by
$$d = e^a\wedge\nabla_{X_a} $$
and
$$d^* = -X^a\hook\nabla_{X_a} \,. $$

The Clifford algebra of each cotangent space is generated by the basis
1-forms. It may be identified with the vector space of exterior forms 
with  Clifford multiplication related to the exterior and interior
derivatives by
$$A\phi = A\wedge\phi +A^\sharp\hook\phi $$
for $A$ a 1-form and $\phi$ an arbitrary form. We will juxtapose symbols
to denote their Clifford  product. Whereas the Clifford product of two
homogeneous forms will be inhomogeneous, the Clifford commutator of a 
2-form with another form will preserve the degree of that form. The Clifford
commutator, denoted $[\quad,\quad]$, is related to the exterior product
by
$$[F,\phi] = -2X_a\hook F\wedge X^a\hook\phi $$
where $F$ is a 2-form and $\phi$ is an arbitrary form. 
The Lie algebra formed by 
the 2-forms under Clifford commutation is the Lie algebra of the Lorentz
group. (In general they form the Lie algebra of the appropriate 
pseudo-orthogonal group.) 
Clifford multiplication by the volume form relates a form to its Hodge
dual. In forming Clifford products it is convenient to 
denote the volume form by $z$, that is
$$z=*1\,. \eqno(4)
$$
We then have 
$$ *\phi = \phi^\xi z $$
where $\xi$ is the Clifford algebra involution that reverses the order
of products and leaves the 1-form generators fixed.
Since Hodge duality preserves the space of 2-forms, squaring to minus
the identity, the space of complex 2-forms can be decomposed into
self-dual and anti-self-dual subspaces satisfying $*F=iF$ and $*F=-iF$
respectively. It then follows that these sub-spaces form simple ideals
in the 	Clifford commutator Lie algebra.

Forms of even degree form a  sub-algebra  of the Clifford algebra.
The complexified even sub-algebra has two inequivalent irreducible 
representations, the spinor representations. Again we will simply
juxtapose symbols to denote the Clifford action on a spinor. So if
$\psi$ is a spinor and $\omega$ any element of the Clifford
algebra we write $\omega\psi$ to denote the Clifford action of $\omega$
on $\psi$. 
The eigenvalues of the volume form label the inequivalent spinor 
representations.
A spinor $\psi$ is even or odd according to whether $iz\psi=\psi$
or $iz\psi=-\psi$. We have  a spin-invariant symplectic product,
which we will denote by a bracket $(\quad,\quad)$, 
that is block diagonal on the inequivalent spinor spaces.
 This product will be chosen such that
$$(u,\omega v) = (\omega^\xi u,v)\,. \eqno(5)$$
This spinor
product gives an isomorphism with the space of dual spinors. We let
$\bar u$ denote the dual spinor such that
$$\bar u(v)=(u,v)\,. $$
Since tensor products of spinors and their duals  are linear transformations
on the space of spinors, 
$$(u\otimes\bar v)w = (v,w)u\,, $$
we may naturally identify such tensors with elements
of the Clifford algebra. 
Under Clifford multiplication by an arbitrary form $\phi$ we have
$$\eqalignno{
\phi(u\otimes\bar v) &= \phi u\otimes\bar v \cr
(u\otimes\bar v)\phi &= u\otimes\overline{\phi^\xi v}\,.\cr}$$
Under the involution $\xi$ we have
$$(u\otimes\bar v)^\xi = -v\otimes\bar u \,. \eqno(6)$$
The parity of the spinors determines that of their tensor product.
For example, if $u$ and $v$ lie in the same spinor space then 
$u\otimes \bar v$ 
is an even form, as the spinor product is zero on spinors of different parity.
Equation (6) shows that the symmetry properties of the tensor product
determine the eigenvalue of $\xi$. For example, the symmetric combination 
$u\otimes\bar v+v\otimes\bar u$ is necessarily odd under $\xi$.
So if both spinors lie in the same spinor space this combination is
then a 2-form.
Moreover, this 2-form will be self-dual if both $u$ and $v$
are even. 
We may expand 
a tensor product of spinors  into $p$-form components as
$$u\otimes\bar v = {1\over4}(v,u) + {1\over4}(v,e_au)e^a
-{1\over8}(v,e_{ab}u)e^{ab} + {1\over4}(v,e_azu)e^az
-{1\over4}(v,zu)z\,. \eqno(7)$$
Here we use the abbreviated notation $e^{ab}$ to denote $e^a\wedge e^b$.

Since, in the Lorentzian case, the irreducible representations of the
complexified Clifford algebra are the complexifications of those of the
real Clifford algebra, we may choose the 
conjugate linear charge conjugation operator such that it is related to 
complex conjugation by
$$\eqalignno{
(u,v)^* &=(u^c,v^c) \cr
\noalign{\hbox{and}}
(u^c)^c&=u \,.\cr}$$

In general we follow the conventions of [14].


\noindent{\bf III. Algebraically-Special Space-times.}
\vskip 0.5true cm
In this section we summarise the Petrov classification of the curvature
tensor. In particular, we will state the condition that a space-time be
algebraically-special in a form  convenient to us. 

In a four-dimensional Lorentzian	 
space-time the Hodge dual map squares to
minus one when acting on 2-forms.  With the Hodge dual as complex 
structure the space of 
2-forms may be
regarded  as a three-dimensional complex space. 
The Lorentzian metric induces a metric on the space of 2-forms. By using
the complex structure of Hodge duality this metric defines a complex
Euclidean structure  on the space of 2-forms. 
The curvature tensor
can be thought of as a map on the space of 2-forms in such a way that, in
an Einstein space, it
commutes with Hodge duality and  may thus be regarded as a complex linear
map. It is also self-adjoint with respect to the complex Euclidean structure.
The Petrov classification scheme classifies the Jordan canonical forms
of the curvature tensor. Details can be found in [15] and [16].

The metric tensor induces a metric on the space of $p$-forms. 
If $\phi$ and $\psi$ are $p$-forms then their scalar product $\phi\cdot\psi$
is defined by
$$\phi\wedge *\psi = \phi\cdot\psi *1\,, $$
which becomes
$$\phi\cdot\psi = 
{1\over2}X_a\hook X_b\hook\phi X^a\hook X^b\hook\psi$$
when $\phi$ and $\psi$ are 2-forms.
We may use this metric to regard the tensor product of two 2-forms as
an endomorphism on the space of 2-forms;
$$(\phi\otimes\psi)(F) = (\psi\cdot F) \phi \,.$$
Clearly, symmetric tensor products correspond to self-adjoint operators,
and thus more generally so do tensors with `pairwise interchange symmetry'. 
Those tensors that are double-self-dual correspond to endomorphisms on
the space of self-dual (or anti-self-dual) 2-forms. 

The curvature tensor may be regarded as an endomorphism on 2-forms by
using the metric to relate it to a totally covariant tensor,
$$R = 2R^{ab}\otimes e_{ba} $$
where $\{R^{ab}\}$ are the curvature 2-forms. The double-self-dual part
of the curvature tensor $R^+$ is related to the conformal tensor $C$ and the
curvature scalar $\cal R$ by
$$\eqalignno{
     R^+ &= C -{1\over6}{\cal R}e_{ab}\otimes e^{ab} \cr
	 &= C -{1\over3}{\cal R} I \cr }$$
where $I$ is the identity map on 2-forms. The conformal tensor $C$ can
be expressed in terms of the conformal 2-forms as
$$C = 2C^{ab}\otimes e_{ba} \,.$$
Thus acting on an arbitrary 2-form $\phi$ we have
$$C\phi = 2X_a\hook X_b\hook\phi C^{ab}\,. \eqno(8)$$

The Petrov type of a space-time is determined by the number of 
eigenvectors and eigenvalues of the
conformal tensor when acting in this way on the space of self-dual 2-forms. 
An algebraically-general conformal tensor has three linearly-independent
eigenvectors with distinct eigenvalues, all other cases being classed as
algebraically-special.

Any self-dual 2-form $\phi$ that is null is also decomposable, and hence
has one independent real null eigenvector $K$, $K\hook\phi=0$.
The principal null directions of the conformal tensor correspond to null
self-dual 2-forms that satisfy $\phi\wedge C\phi=0$. (See [15].)
This is clearly satisfied by any null eigenform of $C$. Such eigenforms
correspond to repeated principal null directions. 
An algebraically-special
space-time may be characterised	as one admitting a null eigenform of 
the conformal tensor. The only space-times that admit two  independent 
null eigenforms are Petrov type $D$. In this case the two independent null 
eigenforms have the  same eigenvalue.


\noindent{\bf IV. Conformal Killing-Yano Tensors.}
\vskip 0.3true cm
Killing tensors were introduced as tensors which obey generalisations
of Killing's equation, with conformal Killing tensors obeying analogues of the
vector conformal Killing equation. 
One generalisation was to replace the vector with a
totally symmetric tensor, whilst Yano [17]
extended Killing's equation to a
totally anti-symmetric tensor. We shall (as is now common) refer to the 
latter as Killing-Yano tensors, reserving the term Killing tensor for a 
totally symmetric tensor. 

Killing's equation expresses the condition that a vector field generate
an isometry in terms of the symmetrised covariant derivative of the
vector field. The Killing tensor and Killing-Yano tensor equations are
also usually expressed in terms of the symmetrised covariant derivative. 
However,  the efficiency of the exterior calculus 
enables the Killing-Yano equation to be written much more  succinctly 
in terms of the anti-symmetrised covariant derivative.
If $K$ is any vector field then the covariant derivative $\nabla K^\flat$
can be decomposed into symmetric and skew parts. The symmetric part can be
further decomposed into a trace-free part and the trace.
The symmetrised covariant derivative is related to the Lie derivative of the
metric tensor. In four dimensions we have
$$\nabla_XK^\flat ={1\over2}X\hook dK^\flat -{1\over4}X^\flat d^*K^\flat
+{1\over2}\left\{{\cal L}_Kg -{1\over4}\Tr({\cal L}_Kg)g\right\}(X)$$
where ${\cal L}_K$ denotes the Lie derivative and $\Tr$ the trace.
(As is usual, we use the metric tensor to regard any degree two tensor
as a linear map on vector fields.)
So the conformal-Killing equation can be written (in four dimensions) as
$$\nabla_XK^\flat ={1\over2}X\hook dK^\flat -{1\over4}X^\flat d^*K^\flat 
\quad\forall X\,. \eqno(9)$$
If in addition $d^*K^\flat=0$ then $K$ is a Killing vector.
Notice that (9) implies that
$$e^a\wedge\del a K^\flat ={1\over2}e^a\wedge X_a\hook dK^\flat 
= dK^\flat\,.$$
Since $e^a\wedge\del a=d$ the coefficient of $X\hook dK^\flat$ is just such
that we cannot conclude that $dK^\flat=0$. Similarly (9) implies that
$$X^a\hook\del aK^\flat =-{1\over4}X_a\hook(e^a  d^*K^\flat )
=-d^*K^\flat$$
in four dimensions. Thus the coefficient of $X^\flat d^*K^\flat$ is just such
that we cannot conclude from (9) that $d^*K^\flat=0$. This observation
suggests how the conformal Killing equation (9) can be generalised to 
forms of higher degree: the covariant derivative is related to the
exterior derivative and the co-derivative with the coefficients chosen such that
we do not automatically have the form closed or co-closed.
If $\omega$ is a 2-form then this generalisation gives the 
 equation 
$$3\nabla_X\omega = X\hook d\omega -X^\flat\wedge d^*\omega
\quad\forall X\,.\eqno(10)$$
From this we can use 
$\nabla_X\omega(Y,Z) ={1\over2}Z\hook Y\hook\nabla_X\omega$
to show that $\omega$ must also satisfy
$$\eqalignno{\nabla_Y\omega (X,Z) +\nabla_Z\omega(X,Y) 
&={1\over3}d^*\omega(X)g(Y,Z) \cr
&\quad - {1\over6}d^*\omega (Y)g(Z,X) 
- {1\over6}d^*\omega(Z)g(X,Y) \qquad\forall X,Y,Z\,,&(11)\cr} $$
which is Tachibana's conformal generalisation of Yano's Killing equation
[2]. Now taking $X=X_a$ and $Y=X_b$ in equation (11), and multiplying
both sides by $e^{ab}$, we recover equation (10). Hence (10) and (11)
are equivalent and we may adopt (10) as the conformal
Killing-Yano (CKY) equation.  

The CKY equation is invariant under Hodge duality. Equation (1) shows
that
$$\eqalignno{
*(X^\flat\wedge d^*\omega) &= -X\hook*d^*\omega\cr
&= -X\hook **d*\omega\quad\hbox{by (3)}\cr
&= -X\hook d*\omega\quad\hbox{by (2).}\cr
\noalign{Equivalently}
*\left(X\hook d\omega\right) &=(X^\flat\wedge *d\omega)\,.\cr}$$
Since $\nabla_X*=*\nabla_X$ it follows that $*\omega$ is a CKY tensor if 
$\omega$ is. Thus any solution to the CKY equation can be decomposed into
self-dual and anti-self-dual CKY tensors.

The CKY equation also often appears in yet another guise. Elements of the
Clifford algebra are naturally identified with tensors on the space of 
spinors. More generally, tensor products of exterior forms may be regarded
as higher degree tensors on the spinor spaces, and so any equation
for an exterior form can also be written in spinor notation.
The CKY equation can be written as
$$\eqalignno{
\Omega &=0 \cr
\noalign{\hbox{where}}
\Omega &=\Omega_a\otimes e^a \cr
\noalign{\hbox{with}}
\Omega_a &= 3\nabla_{X_a}\omega -X_a\hook d\omega +e_a\wedge d^*\omega\,.
\cr}$$
The tensor $\Omega$ can be thought of as a tensor acting on three even
spinors $u,v$ and $w$, and one odd spinor $\alpha$ by
$$\Omega(u,v,w,\alpha) = (u,\Omega_av)(\alpha,e^aw)\,.$$
Since the spin-invariant product satisfies (5) we have
$$\eqalignno{
(u,\Omega_av) &= -(\Omega_au,v)\qquad\hbox{for $\Omega_a$ 2-forms,}\cr
&= (v,\Omega_au)\qquad\hbox{since the product is symplectic.}\cr}$$
Thus $\Omega$ is automatically symmetric in the first two spinors.
It will be totally symmetric in the three even spinors if it is symmetric 
under interchange of $u$ and $w$ say. Since the space of even spinors is
two-dimensional and the product is symplectic we have the identity
$$(u,v)w +(v,w)u +(w,u)v =0 \qquad\hbox{for all even $u,v,w$.} $$
Thus
$$\eqalignno{\Omega(u,v,w,\alpha) &= (u,\Omega_av)(\alpha,e^aw) \cr
&= -\bigl(\alpha,e^a\left\{(\Omega_av,w)u + (w,u)\Omega_av\right\}\bigr)\cr
&=(v,\Omega_aw)(\alpha,e^au) -(w,u)(\alpha,e^a\Omega_av)\cr
&=\Omega(v,w,u,\alpha) - (w,u)(\alpha,e^a\Omega_av) \cr
&=\Omega(w,v,u,\alpha) -(w,u)(\alpha,e^a\Omega_av)\,.\cr}$$
Now
$$e^a\Omega_a = e^a\wedge\Omega_a +X^a\hook\Omega_a\,,$$
and the $\Omega_a$ are such that $e^a\wedge\Omega_a=0$ and 
$X^a\hook\Omega_a=0$.
Thus $\Omega$ is totally symmetric on the three even spinors. 
If $\omega$ is 	self-dual then $\Omega(u,v,w,\alpha)$ will be zero unless
the first three spinors are even and the last is odd. Thus the CKY equation
for a self-dual 2-form can be regarded as a Spin-irreducible tensor-spinor
equation. This spinor equation, equivalent to the CKY equation,
 was introduced in its own right in [18], and is now usually known as 
the Killing spinor equation  [13]. Because 
Killing spinors were introduced separately their correspondence
with CKY tensors has not always been made clear. There is potential
for confusion in that tensors
corresponding to  Killing spinors have also been called Penrose-Floyd 
tensors [19].

In the following section we will consider equations for 2-forms related to 
shear-free congruences. 
The CKY equation can be regarded as a special case of the shear-free
equation.  Integrability conditions for the CKY equation then follow
as special cases of those for the shear-free equation which  will be
derived in the following section. 


{\bf V. Shear-free Equations.}
\vskip 0.3true cm
A congruence of curves may be  specified by a vector field, the tangent 
field. The  congruence is shear-free if the tangent field generates
conformal transformations on its conjugate space (the space of vectors
to which it is orthogonal). Thus the shear-free condition is a generalisation
of the conformal-Killing condition. 
Since the shear-free condition is conformally invariant, the various
shear-free equations that will be given all have a conformal covariance.
If a vector field generates conformal transformations on its conjugate then
so does any vector field proportional to it. Since a  reparametrisation 
of the congruence corresponds to a scaling of
the tangent field the shear-free condition is a 
reparametrisation-invariant property of the
congruence. 
The condition that a congruence be shear-free 
can be formulated 
as a `gauged' conformal Killing equation, where the connection terms 
ensure covariance under a scaling of the vector field. Thus a vector field
$K$ corresponds to a shear-free congruence if it satisfies the 
equation [20]
$$\nablahat_XK^\flat = {1\over2}X\hook\dhat K^\flat -{1\over4}X^\flat
\codhat K^\flat\qquad\forall X. \eqno(12)$$
Here $\nablahat$ is a scaling-covariant derivative,
$$\nablahat_X K^\flat =\nabla_X K^\flat + 2qA(X)K^\flat $$
for some real 1-form $A$. (The factor of 2 and the constant $q$
will be convenient later.)
The gauged exterior derivative $\dhat$ and
co-derivative $\codhat$ are related to $\nablahat$ by
$$\eqalignno{
\dhat &= e^a\wedge\nablahat_{X_a} \cr
\codhat &= -X^a\hook\nablahat_{X_a}\,. \cr}$$
(Equation (12) has the numerical coefficients chosen
for four dimensions, although it is easily generalised to arbitrary 
dimensions.) 
Throughout we shall write various shear-free equations in terms of a
`gauged' covariant derivative. However, it must be remembered that
the form $A$, playing the role of connection, is not some given background
field, but depends upon the vector field $K$. 
When $K$ is non-null then $A$ can be expressed in terms of $K$, whereas in
the null case only certain components can be expressed in terms of 
$K$ [20].

If $K$ is null (and real)
then it may be related to an even spinor $u$ by
$$K^\flat = (iu^c,e^au)e_a\,.$$
Clearly the correspondence between $K$ and $u$ is not one-to-one, there
being a $U(1)$ freedom in the choice of $u$. The shear-free condition
(12) for $K$ is then equivalent to the following equation for $u$:
$$\eqalignno{
\nablahat_X u -{1\over4}X^\flat\Dhat u &=0\qquad\forall X&(13)\cr
\noalign{\hbox{where}}
\nablahat_X u &=\nabla_Xu +q{\cal A}(X)u &{(14)}\cr	
\noalign{\hbox{where $\cal A$ is a complex 1-form whose real part is
$A$, and $\Dhat$ is the Dirac operator}}
\Dhat &= e^a\nablahat_{X_a}\,. \cr
}$$
The shear-free spinor equation (13) is a $\Cstar$-covariant twistor equation,
where $\Cstar$ is the group of non-zero complex numbers. 
The $U(1)$ part of the covariance stems from the projective relationship
between $u$ and $K$, whilst the scaling part of the covariance is related 
to the reparametrisation-invariance of the shear-free condition.
In the same way that we showed the spinorial correspondence of the CKY
equation in the previous section, we may regard (13) as an equation for
a spin tensor acting symmetrically on two even  spinors and one odd spinor.
Written thus the shear-free spinor equation was obtained by 
Sommers [21].

A (real) null vector field $K$ may  
be put into correspondence with a self-dual 
decomposable 2-form $\phi$ by the relation
$$K^\flat\phi =0\,. $$
A given $K$ only determines $\phi$ up to a complex scaling. 
The shear-free condition for $K$ gives rise to an equation for $\phi$.
In fact Robinson's theorem [3] shows that $\phi$ is proportional to a
closed (and hence, since it is self-dual,  co-closed) 2-form; that is, a 
Maxwell solution. It will be convenient for us
to state the shear-free condition
for $\phi$ differently. In terms of the 
even spinor $u$ representing $K$ we may choose 
$$\phi = u\otimes \bar u\,. \eqno(15) $$
The shear-free condition for $u$ then  translates to an equation for $\phi$. 
It is simplest to obtain the  corresponding equation for $\phi$ written 
in terms  of a Clifford commutator, $[\quad,\quad]$. 
If we write $\phi$ as in (15) then
we can show that  $u$ satisfies (13) if and only if
$$\nablahat_X\phi -{1\over4}[X^\flat e^a,\nablahat_{X_a}\phi] =0 \eqno(16)
$$
where
$$\nablahat_X\phi =\nabla_X\phi +2q{\cal A}(X)\phi\,.\eqno(17) $$
Since	the Clifford commutator term can be written as
$$ {1\over2}[X^\flat e^a,\nablahat_{X_a}\phi]  =  X\hook\dhat\phi
-X^\flat\wedge\codhat\phi -\nablahat_X\phi $$
we see that the shear-free equation (13) is equivalent to the gauged
conformal Killing-Yano equation
$$3\nablahat_X\phi = X\hook\dhat\phi - X^\flat\wedge\codhat\phi\,.\eqno(16')$$
(We reiterate that the form $\cal A$ depends upon the 2-form $\phi$: in 
particular, should there be two shear-free 2-forms then, in general,
the `gauge terms' occurring in each equation will be different.)
Thus as an alternative to Robinson's theorem we can state
that a decomposable self-dual 2-form corresponds to a  shear-free 
null congruence if and only if it satisfies the $\Cstar$-gauged conformal
Killing-Yano equation. One can show that null solutions to ($16'$) can be
scaled to produce Maxwell fields, and vice versa, and so indeed this statement
is equivalent to Robinson's theorem. 
Previously Dietz and R\"udiger [4] investigated a 
generalisation of Robinson's theorem. They considered non-decomposable
2-forms corresponding to two independent null congruences. They showed
that both these congruences are shear-free if and only if the self-dual
2-form satisfies an 
equation equivalent to ($16'$). 
Although they showed that their equation was a generalisation of the CKY
equation they did not interpret the extra terms as gauge terms. Moreover,
they only considered non-decomposable solutions to the equation. 
In fact it is rather nice to have a single equation for a 2-form that
characterises any eigenvectors as being shear-free, whether there be one
or two independent real eigenvectors.

The shear-free equations can be differentiated to obtain integrability
conditions relating second derivatives to curvature terms. 
We shall make use of these later on. 
Firstly we 
consider  the shear-free spinor equation. 
Differentiating (13) introduces the  curvature operator 
$\hat R(X,Y)$ of $\nablahat$. Since 
$\nablahat$ is related to $\nabla$ by (14) the curvature operators are
related by
$$\hat R(X,Y)u	= R(X,Y)u -qX\hook Y\hook {\cal F} u \eqno(18)$$
where $\cal F$ is the $\Cstar$ curvature,
$${\cal F} = d{\cal A} \,. $$
Since
$$R(X,Y)u = {1\over2}e^a(X)e^b(Y)R_{ab}u \eqno(19)$$
where $R_{ab}$ are the curvature 2-forms, equation (13) has the 
integrability condition
$$R_{ab}u +2qX_b\hook X_a\hook{\cal F}u -{1\over2}(e_b\nablahat_{X_a} 
-e_a\nablahat_{X_b})\Dhat u =0 \,. \eqno(20)$$
 Multiplying by $e_a$ gives
$$P_bu -2qX_b\hook{\cal F}u +\nablahat_{X_b}\Dhat u +{1\over2}e_b\Dhat^2u
=0 \eqno(21)$$
where $P_b$ are the Ricci 1-forms [14]. Multiplying this by $e^b$ produces
$$ \Dhat^2u =  {4q\over3}{\cal F}u -{1\over3}{\cal R}u \eqno(22)$$
where $\cal R$ is the curvature scalar.
A Laplacian on spinors is given by the trace of the Hessian,
$$\hat\nabla^2 = \nablahat_{X^a}\nablahat_{X_a} 
-\nablahat_{\nabla_{X_a}X^a}\,.$$ 
This is related to the square of the Dirac operator
by curvature terms:
$$\Dhat^2 u = 
\hat\nabla^2 u + {1\over2}e^{ab}\hat R(X_a,X_b)u \,. \eqno(23)$$
By (18) and (19) we have 
$$e^{ab}\hat R(X_a,X_b)u = -{1\over2}{\cal R}u +2q{\cal F}u $$
and so (22) can be written as
$$\hat\nabla^2 u
= {q\over3}{\cal F}u -{1\over12}{\cal R}u \,. \eqno(\hbox{$22'$}) $$

Since the conformal 2-forms $C_{ab}$ are given by
$$C_{ab} =  R_{ab} -{1\over2}(P_a\wedge e_b -P_b\wedge e_a)
+{1\over6}{\cal R}e_a\wedge e_b        $$
we may use (20), (21) and (22) to obtain the integrability condition
$$C_{ab}u + q{\Gamma}_{ab}u = 0 \eqno(24) $$
where
$$\Gamma_{ab} = {1\over6}e_{ba}{\cal F} +{1\over2}{\cal F}e_{ba}\,.
\eqno(25) $$

Now we look at the integrability conditions for the decomposable self-dual
2-form describing the shear-free congruence. If $\phi$ is related to $u$
by (15) then we may use the integrability condition (24) of (13) to obtain
an analogous integrability condition of ($16'$). If we define
$ C\phi$ by (8) then from (15) and (7)
$$\eqalignno{
C\phi &= -{1\over2}(u,e_{ab}u)C^{ba}  \cr
&= -{1\over2}(u,C_{ab}u)e^{ba} \cr }$$
by the `pairwise symmetry' of the conformal tensor. Now we may use (24) to
obtain
$$\eqalignno{
C\phi &= {q\over2}(u,\Gamma_{ab}u)e^{ba}  \cr
&= -{q\over4}(u,(\Gamma_{ab}-\Gamma_{ab}{}^\xi)u)e^{ab}\,. \cr
&= -{q\over12}(u,[{\cal F},e_{ba}]u)e^{ab} \cr
\noalign{\hbox{from (25),}}
&= -{q\over12}(u,e_{ab}u)[{\cal F},e^{ab}]\cr
\noalign{\hbox{due to the `pairwise anti-symmetry' of 
$X_p\hook X_q\hook[{\cal F},e_{ba}]$,}}
&= {2q\over3}[{\cal F},\phi] &(26)\cr
\noalign{\hbox{from (7).}}
}$$

In analogy with (23) the trace of the Hessian is related to the 
`gauged' Laplace-Beltrami operator $\hat\triangle$ by
$$\eqalignno{
\hat\triangle\phi &=
\hat\nabla^2\phi + e^b\wedge X^a\hook\hat R(X_b,X_a)\phi \cr
&= \hat\nabla^2\phi + {1\over4}[e^{ba}, \hat R(X_b,X_a)\phi] \cr }$$
where 
$$\hat\triangle \phi = -(\dhat\codhat +\codhat\dhat)\phi \,. $$
Since $\nablahat$ is related to $\nabla$ by (17) the curvatures are related
by
$$\eqalignno{
\hat R(X,Y)\phi &= R(X,Y)\phi -2qX\hook Y\hook{\cal F}\phi \cr
\noalign{\hbox{and hence}}
[e^{ba},\hat R(X_b,X_a)\phi] &=	[e^{ba}, R(X_b,X_a)\phi] 
                                   +4q[{\cal F},\phi] \cr
&= -2C\phi -{4\over3}{\cal R}\phi +4q[{\cal F},\phi]\,. & \cr}$$
Thus $\hat\triangle$ is related to $\hat\nabla^2$ by
$$\hat\triangle\phi =\hat\nabla^2\phi -{1\over2}C\phi -{1\over3}{\cal R}\phi
+q[{\cal F},\phi]\,.\eqno(27)$$

By differentiating ($16'$) we obtain
$$3\hat\nabla^2 \phi = \hat\triangle\phi\,, $$
and so (26) gives the integrability condition
$$\eqalignno{
\hat\nabla^2\phi &={q\over3}[{\cal F},\phi] -{1\over6}{\cal R}\phi&(28)\cr
\noalign{\hbox{or}}
\hat\triangle\phi +{1\over2}{\cal R}\phi &= q[{\cal F},\phi] \,,&(\hbox{$28'$})\cr
}$$
where the left-hand side is the conformally covariant (gauged) 
Laplace-Beltrami operator.

The integrability condition (26) shows that $\phi\wedge C\phi=0$
and thus the null $\phi$ corresponds to a principal direction. In a 
Ricci-flat space-time the
Goldberg-Sachs theorem makes the stronger statement that a 
shear-free null congruence must correspond to a {\sl repeated} principal null
direction, 
and conversely  any repeated principal null
direction must correspond to a shear-free congruence [22, 23]. 
More generally necessary and sufficient conditions for a shear-free 
congruence to correspond to a repeated principal null direction are
given by the generalised Goldberg-Sachs theorem [24].
So in the generalised Goldberg-Sachs class of space-times 
$\phi$ must be an eigenform of the conformal tensor. Thus from (26)
the commutator of $\cal F$ and $\phi$ must be proportional to $\phi$,
and so from (15) we see that the spinor $u$ must be an eigenvector of
$\cal F$. Thus 
$$q{\cal F}u = 3\mu u\,, \eqno(29)$$
for some eigenfunction $\mu$, and
$$q[{\cal F},\phi] = 6\mu\phi\,. \eqno(30)$$
The integrability condition (22) for the spinor becomes
$$\eqalignno{
\hat D^2 u &=4\mu u -{1\over3}{\cal R}u & (31a) \cr
\hat\nabla^2 u &= \mu u -{1\over12}{\cal R}u \,,&(31b)\cr}$$
whilst (28) becomes
$$\eqalignno{
\hat\nabla^2\phi &= 2\mu\phi -{1\over6}{\cal R}\phi & (32a) \cr
\hat\triangle\phi +{1\over2}{\cal R}\phi &= 6\mu\phi\,, & (32b) }$$
and (26) becomes
$$C\phi = 4\mu\phi\,. \eqno(33) $$

We can obtain integrability conditions for the CKY equation using the results
that we have for shear-free equations. 
In the case of a null self-dual
CKY tensor  we 
simply have the conditions above 
where the `gauge' terms are zero. 
In this case the corresponding shear-free spinor equation reduces to the
twistor equation.
From (26) we see that the CKY tensor 
is a null eigenvector of $C$ with eigenvalue zero. 
In fact (24) shows that any self-dual 2-form $\phi$
that has $u$ as eigenspinor satisfies $C\phi=0$. Therefore $C$ must
have two linearly-independent eigenvectors, and so the space-time must
be type $N$ or conformally-flat.

If  $\omega$ is a non-null self-dual 2-form then we may write $\omega$
in terms of a pair of even spinors $u_1$ and $u_2$ as
$$\omega = {1\over2}(u_1\otimes \bar u_2 + u_2\otimes\bar u_1) \,.
\eqno(34)$$
In the same way that we showed that the shear-free spinor equation (13)
led to the `gauged' conformal Killing-Yano equation  (16) we may show that
the non-null $\omega$ satisfies the CKY equation if and only if the
spinors $\{u_i\}$ both satisfy a shear-free equation, with the 
spinors having opposite $\Cstar$ `charges'. 
(Dietz and R\"udiger showed that the 2-form $\omega$ satisfies a gauged CKY
equation if and only if the spinors $\{u_i\}$  satisfy the shear-free
equation [4]. In general the $\Cstar$ `charge' of $\omega$ is the sum of those
of $u_1$ and $u_2$. So we have here just a special case when $\omega$ is a CKY
tensor.)

We may use the integrability conditions for the spinor equations to
obtain integrability conditions for $\omega$, just as we did in the 
shear-free  case. We have
$$\eqalignno{
C\omega &= -{1\over2}(u_2,e_{ab}u_1)C^{ba} \cr
&= - {1\over2}(u_2,C_{ab}u_1)e^{ba} \cr
&= -{1\over2}(u_1,C_{ab}u_2)e^{ba}\,. \cr}$$

The integrability condition (24) for $u_1$ gives
$$ C\omega = {q_1\over2}(u_2,\Gamma_{ab}u_1)e^{ba}\,, \eqno(35) $$
whilst that for $u_2$ gives
$$\eqalignno{
C\omega &= {q_2\over2}(u_1,\Gamma_{ab}u_2)e^{ba} \cr
        &= {q_1\over2}(u_2,\Gamma_{ab}{}^\xi u_1)e^{ba}\cr }$$
since $q_2=-q_1$. 

Subtracting this from (35) shows that
$$\bigl(u_2,(\Gamma_{ab}-{\Gamma_{ab}}^\xi)u_1\bigr)e^{ba} =0\,.$$
Repeating the steps that lead to (26) then shows that we must have
$$[{\cal F},\omega]=0\,. \eqno(36)$$
From this we can conclude that the self-dual part of $\cal F$ is proportional
to $\omega$. Since $u_1$ and $u_2$ are eigenvectors of $\omega$, and hence 
$\cal F$, there must be some function $\mu$ such that
$$q_i{\cal F}u_i = 3\mu u_i \quad\hbox{(no sum.)}\eqno(\hbox{$29'$})$$

Both of the  $\{u_i\}$  satisfy the integrability conditions ($31a$) 
for the appropriate covariant derivative. From each of the spinors we can
make a null shear-free 2-form:
$$\phi_i = u_i\otimes \bar u_i\quad\hbox{(no sum,)}$$
each satisfying the integrability condition (33).
Thus each of the $\phi_i$  corresponds to a repeated principal
null direction. Thus for a non-null CKY tensor to exist we must have
Petrov type $D$ (or conformally-flat), within the generalised Goldberg-Sachs 
class of space-times.
Each of $\phi_1$ and $\phi_2$ satisfies the integrability conditions ($32a$).

We now return  to the integrability condition (35) for $\omega$. 
If we insert (25) into (35) and use (\hbox{$29'$}) then we have
$$C\omega = - 8\mu\omega\,. \eqno(37) $$
Thus $\phi_1$, $\phi_2$ and $\omega$ form an eigenbasis of self-dual
2-forms under the action of the conformal tensor. 
The CKY tensor $\omega$ also satisfies ($32a$) where the `gauge terms' 
are absent from the Laplace-Beltrami operator. 

We have seen that for a non-null CKY tensor  to exist the space-time is 
necessarily Petrov type $D$. Suppose now that we have Petrov type $D$
with $\phi_1$ and $\phi_2$ null shear-free 2-forms corresponding to
repeated principal null directions. 
We cannot, in general, assume any relation between the gauge terms in
the two different shear-free equations. 
We let $\omega$ be defined by (34) such that $\{\phi_1,\phi_2,\omega\}$
is an eigenbasis for $C$. Since each $\phi_i$ is an eigenvector of $C$
(29) becomes
$$q_i{\cal F}_iu_i = 3\mu_iu_i\quad\hbox{(no sum,)} \eqno(\hbox{$29''$})$$
where now we cannot assume that ${\cal F}_1$ and ${\cal F}_2$ are
related. It then follows that each $\phi_i$ 
is an eigenvector of $C$ with eigenvalue $4\mu_i$,  and since the
eigenvalues are assumed equal we can conclude that
$\mu_1=\mu_2$. Proceeding as we did in (35) we can express $C\omega$
in terms of the integrability conditions for either spinor:
$$\eqalignno{
C\omega &= \left({\mu\over4}(u_2,e_{ba}u_1) 
-{1\over4}(q_1{\cal F}_1u_2,e_{ba}u_1)\right)e^{ba} \cr
&=\left({\mu\over4}(u_2,e_{ba}u_1) 
-{1\over4}(q_2{\cal F}_2u_1,e_{ba}u_2)\right)e^{ba}\,. \cr }$$
Since $\omega$ has been assumed an eigenvector of $C$ we must have
$u_2$ an eigenspinor of ${\cal F}_1$. By taking the product with
$u_1$, which is also an eigenspinor, we see that the eigenvalues
of the two spinors
must be opposite. We can draw similar conclusions for ${\cal F}_2$
and have
$$\eqalignno{
q_1{\cal F}_1u_1 &= 3\mu u_1 \cr
q_1{\cal F}_1u_2 &= -3\mu u_2 \cr
q_2{\cal F}_2u_1 &= -3\mu u_1 \cr
q_2{\cal F}_2u_2 &= 3\mu u_2 \,. \cr }$$
It follows that the self-dual part of $q_1{\cal F}_1+q_2{\cal F}_2$ is
zero. That is, the self-dual part of the curvature of the gauge term
entering into the equation for $\omega$ is zero (as was shown by
Dietz and R\"udiger [4]). It then follows that $\omega$ satisfies
(27) with the $\Cstar$-curvature term on the right-hand side zero.
Hence $\omega$ satisfies ($32a$), and in addition (30).
Thus ($32a$) is satisfied by {\sl any} shear-free 2-form whose eigenvectors
are aligned with repeated principal directions. 

We have seen that in type $D$ space-times the self-dual part of the
$\Cstar$-curvature associated with $\omega$ must vanish.
When the anti-self-dual part is also
zero then $\omega$ can be scaled to a CKY tensor. Penrose and Walker's
result on Killing spinors shows that this can be done in every type $D$ 
Einstein space [18].


\noindent{\bf VI. Debye potentials and symmetry operators for source-free
Maxwell fields.}
\vskip 0.3true cm
Maxwell's equations require that the electromagnetic field 2-form
$F$ be closed. This is automatically satisfied if $F$ is exact, $F=dA$
say, where $A$ is the potential 1-form. In source-free regions $F$
is also required to be co-closed, that is $d^*F=0$. This would be
automatically satisfied if there were a co-potential 3-form $B$ such that
$F=d^*B$. Suppose that a 2-form $Z$ 
satisfies
$$\triangle Z = -d{\cal G} - d^*{\cal W}, \eqno(38)$$ 
where ${\cal G}$ and ${\cal W}$ are
arbitrary forms of degree one and three respectively. Then  
by noting that this can be rewritten as
$$d^*\left({\cal W} - dZ\right) = d\left(d^*Z-{\cal G}\right)\eqno(38')$$
we see that  $Z$ provides us with a 2-form that is both 
exact and co-exact and 
hence  a Maxwell solution.  The 2-form $Z$ is called a Hertz potential 
[25]. On the face of it (38) doesn't offer a promising approach, solving
this equation is no easier than solving the original Maxwell equations.
However, in some  circumstances, with an appropriate choice of the form 
for $Z$, solutions
to (38) can be written in terms of  solutions to a scalar equation; a
Debye potential equation [26, 1 and references therein].  
Cohen and Kegeles [1] applied the Debye potential method to the solution
of Maxwell's equations in curved space-times. They pointed out that 
in algebraically-special space-times there exist privileged 2-forms 
corresponding to the repeated principal null direction. 
Using the Newman-Penrose formalism they explicitly obtained scalar Debye
potentials by aligning the Hertz potential with the repeated principal
null direction: the resulting scalar equation being expressed in the 
adapted null basis.

In this section we show how a Debye potential is obtained 
by choosing $Z$ to be proportional to a shear-free 2-form in an
algebraically-special space-time. 
As we shall explicitly use the shear-free equation, rather than an
adapted basis, the resulting Debye potential equation will be expressed
in a basis-independent form. The scheme becomes more powerful when a
conformal Killing-Yano tensor exists.

Consider a shear-free 2-form $\phi$, satisfying ($16'$). (We allow $\phi$
to be either null or non-null, and in the case in which the $\Cstar$-charge 
is zero then we have a CKY tensor.)
Let $f$ be a scalar field with opposite $\Cstar$-charge so that the 
2-form $f\phi$ is a $\Cstar$-scalar. 
Then the 
Laplacian of $f\phi$ can be expressed as
$$\nabla^2(f\phi) =\hat\nabla^2f\phi +2\hat\nabla_{X_a}f\hat\nabla_{X^a}\phi
+f\hat\nabla^2\phi$$
where $\hat\nabla$ is the $\Cstar$-covariant derivative. We can write this 
in terms of the the $\Cstar$-gradient as
$$\eqalignno{
\nabla^2(f\phi) &=\hat\nabla^2f\phi +2\hat\nabla_{\gradhat f}\phi
+f\hat\nabla^2\phi\,,\cr
\noalign{\hbox{or}}
\triangle(f\phi) &= \hat\nabla^2f\phi +2\hat\nabla_{\gradhat f}\phi
+f\hat\nabla^2\phi -{1\over2}fC\phi -{1\over3}f{\cal R}\phi &(39) \cr}$$
by (27).
Now, from the shear-free equation ($16'$) we have
$$\eqalignno{
3\hat\nabla_{\gradhat f}\phi &= \gradhat f\hook\hat d\phi -\hat df\wedge
\hat d^*\phi \cr
&=X^a\hook\left(\hat\nabla_{X_a}(f\hat d\phi) 
-f\hat\nabla_{X_a}\hat d\phi\right)
-d(f\hat d^*\phi) +f\hat d\hat d^*\phi \cr
&= -d^*(f\hat d\phi) +f\hat d^*\hat d\phi 
-d(f\hat d^*\phi)
+f\hat d\hat d^*\phi \cr
&= -d^*(f\hat d\phi)     - d(f\hat d^*\phi)
-f\hat\triangle\phi\,.\cr}$$
Inserting this into (39) gives
$$\triangle(f\phi) =\hat\nabla^2f \phi -{2\over3}d^*\left(f\hat d\phi\right)
-{2\over3}d\left(f\hat d^*\phi\right) 
+f\left(\hat\nabla^2\phi -{2\over3}\hat\triangle\phi\right)
-{1\over2}fC\phi -{1\over3}f{\cal R}\phi \,.$$
When $\phi$ is self-dual with its eigenvectors  repeated principal 
null directions this becomes, by ($32a$),  
$$
\triangle(f\phi) = \left(\hat\nabla^2f-{1\over6}{\cal R}f\right)\phi 
- \lambda_\phi f\phi
-{2\over3}d^*\left(f\hat d\phi\right) -{2\over3}d\left(f\hat d^*\phi\right)
$$
where 
$$2\mu\phi +{1\over2}C\phi =\lambda_\phi\phi\,. \eqno(40)$$ 
It is convenient  to identify $\lambda_\phi$ as an eigenvalue,
$$C\phi  = n^2\lambda_\phi\phi\,, \eqno(\hbox{$40'$})$$
where $n$ is the number of (real) eigenvectors of $\phi$. Thus when $\phi$
is null $n=1$ and $\lambda_\phi$ is just the eigenvalue of $C$ corresponding
to $\phi$, whereas
in the  non-null case $n=2$ and $\lambda_\phi$ is a 
quarter of the eigenvalue of $C$. 
So if the scalar field satisfies the equation
$$\hat\nabla^2f - {1\over6}{\cal R}f = \lambda_\phi f\eqno(41)$$
then    we can relate an exact form to a co-exact one:
$$d\left(d^*(f\phi) -{2\over3}f\hat d^*\phi\right) =
d^*\left({2\over3}f\hat d\phi -d(f\phi)\right) \,.$$
Thus out of the 2-form $\phi$ and the scalar $f$ we have a source-free
Maxwell solution,
$$\eqalignno{
F^\prime(f,\phi) &= d\left(d^*(f\phi) 
       -{2\over3}f\hat d^*\phi\right) &(42a) \cr
     &=d^*\left({2\over3}f\hat d\phi -d(f\phi)\right) \,. &(42b)\cr}$$
Notice how with this approach, unlike that of Cohen and Kegeles, 
we do not need to choose the gauge terms ${\cal G}$ and ${\cal W}$
by inspection, rather they 
appear naturally and are given explicitly in term of $\phi$.
Using (3) and the fact that $\hat d^*=*\hat d*$, we can see that
$F^\prime(f,\phi)$ is anti-self-dual for self-dual $\phi$.

To apply this scheme to construct a Maxwell solution in a given space-time
it is necessary to solve the scalar equation (41). 
The shear-free 2-form $\phi$ enters into this equation, not only through
the eigenfunction $\lambda_\phi$, but also through the `gauge term'
$\cal A$.  We may expose the gauge terms in the shear-free equation
($16'$) by writing it as
$$\eqalignno{
Y_X\phi &=q[X^\flat\wedge{\cal A},\phi] -4q{\cal A}(X)\phi &(43) \cr
\noalign{\hbox{where we have defined $Y_X$ by}}
Y_X\phi	&=3\nabla_X\phi -X\hook d\phi +X^\flat\wedge d^*\phi\,.&(44)\cr}$$
When there is only one repeated principal null direction then the self-dual
$\phi$	will be null. To use (43) to determine $\cal A$ it is convenient
to pick some adapted basis. (Such a point comes to us all when we have
to actually solve equations.)  Let $\Sigma$  be the maximal totally
isotropic subspace such that $X\in\Sigma\Leftrightarrow X^\flat\phi=0$.
Then for $X$ in $\Sigma$ (43) reduces to
$$Y_X\phi = -2q{\cal A}(X)\phi \quad\forall X\in\Sigma\,.$$
This enables half of the components of $\cal A$ to be found. To compute the
remaining components we may pick a maximal isotropic subspace $\Sigma'$
to complement $\Sigma$. Then let $\psi$ be a null self-dual 2-form such that
$X\in\Sigma'\Longleftrightarrow X^\flat\psi =0$. Then the remaining
components of $\cal A$ can be calculated from (43) which becomes
$$Y_X\phi\cdot\psi = -6q{\cal A}(X)\phi\cdot\psi\quad\forall X\in\Sigma'\,.$$

In a type $D$ space-time there are 
two repeated principal null directions. Thus one can choose 
either of the two corresponding	null shear-free 2-forms as Hertz potentials,
or one can choose the non-null 2-form $\omega$ that has both repeated
principal directions as eigenvectors. 
Thus there are three possible choices of Hertz potential, as was pointed
out by Mustafa and Cohen [27].  They chose to normalise $\omega$ to
have constant length, whereas we have chosen to scale $\omega$ to be CKY
and so their equation for a non-null Debye potential will differs from
ours by exact gauge terms.

In the non-null case the gauge term is determined directly in terms of 
$\omega$. 
In this case (43) gives
$$Y_X\omega\cdot\omega =-4q{\cal A}(X)\omega\cdot\omega\quad\forall X\,.$$
This expression was given by Dietz and R\"udiger [4].
Clearly the scheme simplifies in the case in which we have a CKY tensor,
for then the gauge terms are absent from the Debye potential equation.

Equation ($42a$) enables a Maxwell field to be constructed from a shear-free
2-form and a scalar Debye potential satisfying (41). It turns out that we can
conversely take a shear-free 2-form and a Maxwell field and construct
a scalar Debye potential.  If 
$f(F,\phi)=F\cdot\phi$ then the $\Cstar$-covariant Laplacian is given
by 
$$\hat\nabla^2f(F,\phi)
 = \nabla^2F\cdot\phi + 2\nabla_{X_a}F\cdot\hat\nabla_{X^a}\phi
+F\cdot\hat\nabla^2\phi\,,$$
where the $\Cstar$-charge of $f(F,\phi)$ is the same as that of $\phi$.
Since $\phi$ satisfies the shear-free equation ($16'$)
$$\eqalignno{
3\nabla_{X_a}F\cdot\hat\nabla_{X^a}\phi &=
\nabla_{X_a}F\cdot (X^a\hook\hat d\phi) 
-\nabla_{X_a}F\cdot (e^a\wedge\hat d^*\phi) \cr
&= dF\cdot\hat d\phi +d^*F\cdot\hat d^*\phi = 0 \cr }$$
by Maxwell's equations. We may then use (27) to relate $\nabla^2F$
to $\triangle F$, which is zero by Maxwell's equations, to obtain
$$\eqalignno{
\hat\nabla^2f(F,\phi) &= 
\left({1\over2}CF +{1\over3}{\cal R}F\right)\cdot\phi
+F\cdot\hat\nabla^2\phi \cr
&=F\cdot\left({1\over2}C\phi +{1\over3}{\cal R}\phi 
+\hat\nabla^2\phi\right)   \cr
\noalign{\hbox{since $C$ is self-adjoint,}}
&=F\cdot\left({1\over2}C\phi +{1\over6}{\cal R}\phi +2\mu\phi\right)\cr
\noalign{\hbox{by (32a),}}
&= F\cdot(\lambda_\phi\phi +{1\over6}{\cal R}\phi)\,.\cr}$$
So $f(F,\phi)$ satisfies equation  (41): notice, however, that $f(F,\phi)$  
has the {\sl same} `charge' as $\phi$, 
whereas the construction of a Maxwell field from $\phi$ requires a 
Debye potential with opposite charge.

For those space-times that admit a non-null CKY tensor the above gives
a method of mapping any Maxwell field to another. As was shown in section 5,
when we have a non-null CKY tensor $\omega$ we have a pair of null shear-free
2-forms, $\phi_1$ and $\phi_2$, having opposite $\Cstar$-charge. We can
take one of these, $\phi_1$ say, and a given Maxwell field $F$ to construct
a Debye potential $f(F,\phi_1)$. This will then have the appropriate
charge to combine with $\phi_2$ to form a new Maxwell field (also note
that $\lambda_{\phi_1}=\lambda_{\phi_2}$). That is, we
have a symmetry operator ${\cal L}_{\phi_1\phi_2}$, 
mapping between Maxwell fields, defined by
$${\cal L}
_{\phi_1\phi_2}F = F^\prime(F\cdot\phi_1,\phi_2)\,.$$
Interchanging the roles of $\phi_1$ and $\phi_2$ gives another symmetry
operator.
However, by using the shear-free equations for $\phi_1$ and $\phi_2$ it can
be shown that when acting on a Maxwell field $F$, their difference vanishes
and so
$${\cal L}_{\phi_1\phi_2}F={\cal L}_{\phi_2\phi_1}F.$$

We can also use a non-null CKY $\omega$ directly to make
a symmetry operator for Maxwell fields.  The scalar $f(F,\omega)$ is a
Debye potential satisfying an uncharged equation and hence can be
combined again with $\omega$ to produce a Maxwell field.  Hence
$${\cal L}_{\omega\omega}F = F^\prime(F\cdot\omega,\omega)$$
is another symmetry operator. 
Since self-dual and anti-self-dual 2-forms are mutually orthogonal, 
these symmetry operators map only the 
self-dual part of a Maxwell field to an anti-self-dual Maxwell field.
To see the relationship between
these symmetry operators, we will first recast them in terms of a 
higher order generalisation of a CKY 2-form.

The CKY tensor $\omega$ enters quadratically in the symmetry operator 
${\cal L}_{\omega\omega}$. By taking the tensor product of $\omega$ with itself
we obtain a tensor, quadratic in $\omega$, which we may regard as an 
endomorphism on the space
of 2-forms, as we did in section 3. 
In this way we may write ${\cal L}_{\omega\omega}$ in terms of a degree-four tensor
constructed from $\omega$. Let $P^+$ be the operator that projects out the
self-dual part of any 2-form. Then clearly $P^+$ commutes with the Hodge
dual. Since it is self-adjoint with respect to the metric on 2-forms it
corresponds to a pairwise-symmetric tensor. Then out of the self-dual
$\omega$ we construct
the tensor $K$:
$$K =\omega\otimes\omega -{1\over3}(\omega\cdot\omega)P^+ \,.\eqno(45)$$
In section 4 we showed how $\omega$ corresponded to a spin-tensor. In the
same way one can show that $K$ corresponds to a totally symmetric spin 
tensor. The symmetry operator constructed from $\omega$ can be written
in terms of $K$ as
$$\eqalignno{
{\cal L}_{\omega\omega}F &= d\left(d^*(KF) -{2\over5}D^*K(F)\right) &(46a) \cr
&= d^*\left({2\over5}DK(F) -d(KF)\right) &(46b) \cr}$$
where the `exterior derivative' $D$ is defined by
$$DK(G) =d(KG) -e^a\wedge K(\nabla_{X_a}G) $$
for $G$ an arbitrary 2-form. The `co-derivative' $D^*$ is defined 
analogously. 

The only non-zero inner products between a CKY $\omega$ and its 
associated oppositely charged null shear-free 2-forms $\phi_1$ 
and $\phi_2$ are related by
$$\phi_1\cdot\phi_2 = -2\omega\cdot\omega\,.$$
After calculating the action of $K$ on the self-dual
basis $\{\omega,\phi_1,\phi_2\}$ we can use this to see that an 
alternative expression for $K$ is
$$K = {1\over2}\phi_1\otimes\phi_2 + {1\over2}\phi_2\otimes\phi_1
     -{1\over3}\phi_1\cdot\phi_2P^+\,.$$
Then using the shear-free equations for $\phi_1$ and $\phi_2$ 
we can rewrite the right hand side of ($46a$) to show that
$$\eqalignno{{\cal L}_{\omega\omega}F
 &={1\over2}{\cal L}_{\phi_1\phi_2}F
          +{1\over2}{\cal L}_{\phi_2\phi_1}F\cr
 &= {\cal L}_{\phi_1\phi_2}F = {\cal L}_{\phi_2\phi_1}F\,,\cr}$$
since we have already seen that the two terms on the right hand side are 
equal when acting on Maxwell fields.
Hence, as was pointed out by 
Torres del Castillo [11] who wrote down these operators using the
two-component spinor formalism, the various Debye schemes give rise 
only one symmetry operator.

The CKY equation for $\omega$ can be used to obtain an analogous
equation for $K$. The analogy is closest if we write the CKY equation 
in terms of the Clifford commutator as in (16). 
For any 2-form $\phi$ let $L_\phi$ be the operator that maps any 2-form to
the Clifford commutator: $L_\phi(\psi)=[\phi,\psi]$.
Then the CKY equation can be written as
$$\nabla_X\omega -{1\over4}L_{X^\flat\wedge e^a}\nabla_{X_a}\omega =0\,,$$
whilst $K$ satisfies the equation
$$\nabla_XK -{1\over6}[L_{X^\flat\wedge e^a}, \nabla_{X_a}K] =0\,.\eqno(47)$$
Here the bracket denotes the commutator of the operators. 
One can show that
equation (47)  corresponds to a spin-irreducible
spin-tensor equation,  as we showed is the case for the CKY equation.
Just as the CKY equation corresponds to the Killing spinor equation, 
equation (47) corresponds to the 
`4-index Killing spinor' equation.
Kalnins, McLenaghan and Williams 
obtained a symmetry operator for Maxwell fields
from a 4-index Killing spinor [12]. They then obtained a corresponding
tensor equation which they observed was analogous to the CKY equation
as written by Tachibana [2] (equation (11)). 


{\bf VII. Debye potentials and symmetry operators for massless Dirac fields.}
\vskip0.3truecm
In the previous section we showed how we could associate a Debye potential
with a shear-free 2-form, enabling the source-free Maxwell equations to be
solved in terms of solutions to a scalar equation.  In the more special case 
in which there existed a CKY tensor, we showed the relation between the Debye
potential scheme and a symmetry operator 
 constructed from the CKY tensor.
In this section we shall show the
analogues of these constructions for  massless Dirac fields.

Let $u$ be an even shear-free spinor corresponding to a repeated
principle null direction, and let $f$ be a scalar field with opposite 
$\Cstar$-charge. Then the odd spinor $\psi'(f,u)$ given by
$$\eqalignno{
\psi'(f,u) &= \dhat f u + {1\over 2}f\Dhat u &(48)\cr }$$
is a $\Cstar$-scalar. 
The action of the Dirac operator on $\psi'(f,u)$ is
$$\eqalignno{
D\psi'(f,u) &= e^a\left(\nablahat_{X_a}\dhat f u + \dhat f\nablahat_{X_a} u
+ {1\over 2}\nablahat_{X_a} f \Dhat u + {1\over 2}f\nablahat_{X_a} \Dhat u \right)\cr
&= \dhat ^2f u 
- \codhat\dhat f u + e^a\dhat f\nablahat_{X_a} u
+ {1\over 2}\dhat f \Dhat u + {1\over 2}f\Dhat^2 u \cr
&= \hat\nabla^2 f u + \dhat ^2f u + {1\over 2}f\Dhat^2 u
+ 2 \left( \nablahat_{\gradhat f} u - {1\over 4}\dhat f\Dhat u\right)\cr
&=\left(\hat\nabla^2f -{1\over6}{\cal R}f\right)u +2\mu fu +\hat d^2f u \cr
}$$
by the shear-free equation (13) and its integrability condition ($31a$).
If the $\Cstar$-charge
of $u$ is $q$, then the charge of $f$ is $-q$ and we have
$$\eqalignno{
\dhat^2 f &= -qf{\cal F} \cr
\noalign{\hbox{and so}}
\dhat^2 fu &= -3\mu f u\quad\hbox{by (5.34).}\cr}$$
Thus
$$D\psi'(f,u) =\left(\hat\nabla^2 f -{1\over6}{\cal R}f\right)u -\mu f u\,,$$
and so $\psi'(f,u)$ satisfies the massless Dirac equation if $f$ satisfies
the `Debye potential' equation
$$\hat\nabla^2 f -{1\over6}{\cal R}f = {1\over4}\lambda_\phi f \eqno(49)$$
where $\phi =u\otimes \bar u$ 
is an eigenvector of $C$ with eigenvalue  $\lambda_\phi$.
This equation is of the same form as that for the Debye potential used
to construct a Maxwell field from $\phi$. However, here the $\Cstar$-charge 
of the scalar potential	is half that required in the Maxwell case, and the
eigenvalue is also different. In the special case in which the `gauge terms'
are zero the shear-free spinor satisfies the twistor equation. In this case
the construction of a massless Dirac solution from a scalar field satisfying 
the conformally-covariant wave equation is an example of what Penrose
has called `spin raising' [13].

As in the Maxwell case, we may form a scalar potential from a shear-free
spinor and a massless Dirac solution. In the special case in which the
shear-free equation reduces to the twistor equation this corresponds
to Penrose's `spin lowering' [13]. 
Out of the  shear-free spinor $u$ and massless Dirac solution $\psi$
we form a scalar $f(u,\psi)$ from the scalar product $(u,\psi)$,
$$f(u,\psi) = (u, \psi)\,. \eqno(50)$$ 
(Unfortunately the notation doesn't work well for us here. The brackets	
on the left-hand side  denote that the scalar is 
constructed from $u$ and
$\psi$, whereas the  bracket on the right-hand side denotes the symplectic
product of the two spinors.)   The scalar $f(u,\psi)$ has the same
$\Cstar$-charge as $u$, and we may evaluate the gauged Laplacian 
of it by noting that the gauged covariant derivative is compatible with the
spinor product:
$$\hat\nabla^2 f(u,\psi) = 
(\hat\nabla^2 u, \psi) 	+ 2 (\widehat{\nabla}_{X^a}u, \nabla_{X_a}\psi)
+ (u, \nabla^2 \psi) \,. $$
Since $u$ satisfies the shear-free equation (13)
$$\eqalignno{
(\hat\nabla_{X^a}u,\nabla_{X_a}\psi) &= 
{1\over4}(e^a\hat Du,\nabla_{X_a}\psi ) \cr
&={1\over4}(\hat Du,D\psi) =0 \cr}$$
since $\psi$ satisfies the massless Dirac equation. We may then use (23)
to relate the spinor Laplacian to the square of the Dirac operator, 
which gives zero when acting on $\psi$, to give
$$\eqalignno{
(\hat\nabla_{X^a}u,\nabla_{X_a}\psi) 
&= (\hat\nabla^2u+{1\over4}{\cal R}u ,\psi) \cr
&=(\mu u +{1\over6}{\cal R}u, \psi ) \cr
\noalign{\hbox{by the integrability condition ($31b$),}}
&=(\mu+{1\over6}{\cal R})f(u,\psi)\,. \cr}$$
Thus the scalar $f(u,\psi)$ satisfies the `Debye potential' equation
(49). Note, however, that $f(u,\psi)$ has the same `charge' as $u$, opposite
to that required to combine with $u$ to make a massless Dirac solution.

To construct one Dirac solution from another we need a pair of shear-free
spinors with opposite `charges', and this is just the case in which we have
a CKY tensor. In that case we may proceed as in the Maxwell case and define
the symmetry operator ${\cal L}_{u_1u_2}$ by
$$\eqalignno{
{\cal L}_{u_1u_2}\psi  &=\psi'((u_1,\psi),u_2) \cr
         &=\hat d(u_1,\psi)u_2 +{1\over2}(u_1,\psi)\hat D u_2\,.&(51)\cr}$$
By interchanging the two spinors in this construction we could have formed
the operator ${\cal L}_{u_2u_1}$. However, as we shall see, these two
operators are in fact the same when acting on massless Dirac fields.  
Since $\nablahat$ is
compatible with the spinor inner product,
$$\eqalignno{
\dhat (u_1,\psi) 
&= (\nablahat_{X_a} u_1, \psi)e^a +(u_1,\nabla_{X_a}\psi)e^a \cr
&= {1\over 4}(e_a \Dhat u_1, \psi)e^a 
+ (u_1, \nabla_{X_a}\psi)e^a \quad\hbox{by (13).}\cr
\noalign{\hbox{Thus}}
\hat d(u_1,\psi)u_2 &={1\over4}(e^au_2\otimes \overline{e_a\hat Du_1})(\psi)
+(e^au_2\otimes\bar u_1)(\nabla_{X_a}\psi) \cr
&={1\over4} e^a(u_2\otimes\overline{\hat Du_1})e_a\psi +
{1\over2}e^a(u_2\otimes\bar u_1 +u_1\otimes\bar u_2)(\nabla_{X_a}\psi)\cr
&\quad +{1\over2}e^a(u_2\otimes\bar u_1 
-u_1\otimes\bar u_2)(\nabla_{X_a}\psi)\,.&(52)\cr
}$$
If $\phi$ is any $p$-form then 	 $e^a\phi e_a= (4-2p)(-1)^p\phi$. So
if $\phi$ is odd, a sum of a 1-form and a 3-form, this becomes
$e^a\phi e_a =-2\phi^\xi$. We can then use (6) to rewrite the
first term in (52) using
$${1\over4} e^a(u_2\otimes\overline{\hat Du_1})e_a
={1\over2}\hat Du_1\otimes\bar u_2 \,. \eqno(53)$$
The last term in (52) is made up of an anti-symmetric combination of
two spinors of the same parity. Thus this even form is even under $\xi$,
and is thus a sum of a 0-form and a 4-form. These forms commute and
anticommute respectively with the 1-form $e^a$ to enable this last
term to be written in terms of the Dirac operator on $\psi$, which
vanishes since $\psi$ is assumed to satisfy the massless Dirac equation.
Thus we can write the symmetry operator on $\psi$ as
$${\cal L}_{u_1u_2}\psi ={1\over2}(\hat Du_1\otimes \bar u_2
+\hat D u_2\otimes \bar u_1)(\psi)
+{1\over2}e^a(u_2\otimes\bar u_1 +u_1\otimes\bar u_2)(\nabla_{X_a}\psi)\,.$$
In this form the two spinors enter symmetrically. Thus
$${\cal L}_{u_1u_2}\psi ={\cal L}_{u_2u_1}\psi$$
for $\psi$ a massless Dirac solution. 
The symmetric tensor product of the two spinors is the CKY tensor 
$\omega$, in terms of which the symmetry operator can be written 
more concisely.
Differentiating $\omega$, and using the shear-free spinor equation,
produces
$$\nabla_{X_a}\omega ={1\over2}\hat\nabla_{X_a}u_1\otimes\bar u_2
+{1\over4}u_1\otimes \overline{e_a\hat Du_2} 
+{1\over2}\hat\nabla_{X_a}u_2\otimes\bar u_1
+{1\over4}u_2\otimes \overline{e_a\hat Du_1}\,.$$
Clifford multiplication by $e^a$ then shows that
$$d\omega -d^*\omega ={3\over4}(\hat Du_1\otimes\bar u_2
+ \hat Du_2\otimes\bar u_1) \,,$$
by (53), and hence we may express the symmetry operator on $\psi$ as
$${\cal L}_{u_1u_2}\psi = e^a\omega\nabla_{X_a}\psi 
+ {2 \over 3}d\omega\psi - {2 \over 3}d^*\omega\psi \,. $$
This operator, constructed out of a self-dual $\omega$, maps even solutions 
of the massless Dirac equation into odd 
solutions, and annihilates odd spinors.
Clearly we could have taken an anti-self-dual CKY tensor and constructed
a symmetry operator that maps odd Dirac to solutions to even ones. 
So more generally we have a symmetry operator ${\cal L}_\omega$ constructed
out of any CKY tensor $\omega$ (with no assumptions of self-duality)
$${\cal L}_\omega\psi =
e^a\omega\nabla_{X_a}\psi 
+ {2 \over 3}d\omega\psi - {2 \over 3}d^*\omega\psi \,.$$
It can be seen directly from the CKY equation, and its integrability
conditions, that this is indeed a symmetry operator. In fact one can show that
$$\eqalignno{
[D, {\cal L}_\omega] 
&= \left(\omega D - {1\over 3}d\omega + d^*\omega\right)D \,.
}$$
Operators, such as ${\cal L}_\omega$, whose commutator with $D$ is of the 
form $RD$ (where $R$ is another	operator) are called {\sl $R$-commuting}.

Kamran and McLenaghan  [19] have obtained the most general
first-order $R$-commuting operator for the Dirac operator.
They showed that 
the non-trivial terms in this operator are constructed from
conformal Killing-Yano tensors of degree 1, 2 and 3. The operator
constructed from the conformal Killing vector is just the Lie
derivative with the appropriate conformal weight, corresponding to the
well known conformal covariance of the equation. A conformal Killing-Yano
3-form is just the dual of a conformal Killing 1-form. Although
they don't explicitly point it out, the operator that
Kamran and McLenaghan construct from the 3-form is essentially just the
operator formed from the corresponding conformal Killing vector
(the operators differ by a term involving the Dirac operator and an
overall factor of $z$). That part of their operator,  ${\cal K}_\omega$,
constructed from `a conformal 
generalisation of a Penrose-Floyd tensor', 
is obtained from a slight modification to ${\cal L}_\omega$:
$${\cal K}_\omega = z\left({\cal L}_\omega -\omega D\right)$$
where $z$ is the volume form. 
The operator ${\cal K}_\omega$, expressed in terms of a Killing spinor,
was obtained independently by Torres del Castillo [11].
Clearly ${\cal L}_\omega$ and ${\cal K}_\omega$
only differ by a factor of $i$ when acting on massless Dirac solutions,
but the commutator of ${\cal K}_\omega$ with the Dirac operator becomes
$$[D, {\cal K}_\omega] = -{2\over 3}zd^*\omega D \,.$$
The conformal Killing-Yano tensor $\omega$ is a Killing-Yano tensor
when $d^*\omega=0$. Thus in this case the above commutator shows that
${\cal K}_\omega$ is a symmetry operator for the {\sl massive}
Dirac equation.
In this case ${\cal K}_\omega$  
has been interpreted as a generalised total angular momentum operator
by Carter and McLenaghan [28].

\noindent{\bf VIII. Discussion and Conclusions.}
\vskip 0.5true cm
There has been much work done on Debye potential methods for solving
massless field equations, and on the construction of symmetry operators
for these equations in algebraically-special space-times. However, the
relation between different approaches has not always been made clear.
We believe that the basis-independent formalism that we have given here
makes it clearer to see the ingredients	that have gone into the various
constructions. We have explicitly shown the relationship between symmetry
operators constructed from Debye potentials and those given by Kalnins,
McLenaghan and Williams [12] constructed from a 4-index Killing spinor. 

In this paper we have only considered the case of a four-dimensional
Lorentzian space-time. One potential advantage of the approach that we
have adopted is that it should be easier to see which features of the
results translate to different dimensions and signatures. 
The adapted basis of the Newman-Penrose formalism 
is of course optimised for the Lorentzian case, and any 
explicit component expression will become unwieldy in higher dimensions.
Whereas one can always (subject to the usual topological caveats) introduce
spinors	in any number of dimensions, there are many features of the 2-spinor
calculus that are special to four dimensions.
Four dimensions are also rather special for Maxwell's equations:
2-forms are `middle forms'. 
If we regard Maxwell's theory as a gauge theory then the Maxwell forms
will be 2-forms regardless of the dimension. However, the geometrical
relationship between null Maxwell solutions and shear-free null geodesics
is carried over to higher dimensions to a relationship between `middle
forms' and certain foliations [29]. 
Conformal Killing-Yano tensors are of course readily introduced in
any number of dimensions (indeed Tachibana [2, 30] considered 
arbitrary dimensions, although positive-definite signature). 
In four dimensions conformal Killing-Yano 2-forms are the only
non-trivial
generalisation of conformal Killing vectors. The 0-forms and 4-forms
are parallel if they are CKY whilst the 3-form is just the Hodge dual
of a conformal Killing vector. In higher dimensions there are of course
more possibilities. We hope to see which of these tensors can be used
to generate symmetry operators for various massless field equations.

The symmetry operator given by the Debye scheme 
is expressed in terms of a `4-index Killing spinor' that is formed
from the tensor product of the CKY 2-form. However, 
for this to be a symmetry operator we only need
the `4-index Killing spinor' equation 
to be satisfied; it is not necessary that the `4-index Killing spinor'
be a product of `2-index' ones. At this point in the paper we have not
given full details of how our calculations were performed. We hope to 
give a fuller account of `generalised conformal Killing-Yano tensors'
in a more general setting later. 

There are a number of aspects of Debye potentials and symmetry operators
that we have not discussed here. One is the application to higher
spin fields.  In conformally-flat
space-times Debye potentials can readily be extended to include massless 
fields of 
arbitrary spin [13].  A Debye potential
prescription for spin-${3\over2}$ fields in electro-vac space-times,
within the
generalised Goldberg-Sachs class, has been given by Torres del Castillo, 
who has also discussed Debye potentials for spin-2
fields [31, 32, 8].
An important application of symmetry operators, and their relation
to CKY tensors, is the question of separation of variables.  The 
separation constants obtained in this procedure can be given an intrinsic
characterisation in terms of eigenvalues of symmetry operators.
Torres del Castillo [33] has shown how, for Maxwell fields,
the Starobinsky constant is given by the symmetry operator obtained via
Debye potential methods in the Pleba\'nski-Demia\'nski background and 
Silva-Ortigoza [34] 
has presented a similar analysis for the Rarita-Schwinger
(spin-${3\over2}$) equation.  Recently Kalnins, Williams and Miller [35]
have given a detailed account of the separation of variables for
electromagnetic and gravitational perturbations in the Kerr space-time
using Hertz potentials. 

As a final note, we point out that not only have Debye potential methods
been successfully applied to many cosmologically interesting space-times
[36, 37], 
they have also lead to new solutions in the seemingly
unrelated fields of isotropic elastic media [38] and force-free
magnetic fields [39].

\vskip 1true cm
\noindent{\bf Acknowledgment.} IMB thanks R. W. Tucker for kindling his
interest in this area and for hospitality whilst visiting the University
of Lancaster. This research was supported by a Newcastle University
research management committee grant. PRC was supported by an APRA 
scholarship.

\vfill\eject

{\bf References}

\item{[1]}
J~M Cohen and L~S Kegeles.
 Electromagnetic fields in curved spaces: A constructive procedure.
 {\it Physical Review D}, 10(4):1070--1084, 1974.

\item{[2]}
Shun ichi Tachibana.
 On conformal {K}illing tensor in a {R}iemannian space.
 {\it T{\^o}hoku Math. Journ.}, 21:56--64, 1969.

\item{[3]}
I Robinson.
 Null electromagnetic fields.
 {\it Journal of Mathematical Physics}, 2(3):290--291, 1961.

\item{[4]}
W~Dietz and R~R{\"u}diger.
 Shearfree congruences of null geodesics and {K}illing tensors.
 {\it Gravitation and General Relativity}, 12(7):545--562, 1980.

\item{[5]}
L~S Kegeles and J~M Cohen.
 Constructive procedure for perturbations of space-times.
 {\it Physical Review D}, 19(6):1641--1664, 1979.

\item{[6]}
J~M Stewart.
 {H}ertz-{B}romwich-{D}ebye-{W}hittaker-{P}enrose potentials in
  general relativity.
 {\it Proceedings of the Royal Society of London A}, 367:527--538,
  1979.

\item{[7]}
G~F~Torres del Castillo.
 Null Strings and {H}ertz potentials.
 {\it Journal of Mathematical Physics}, 25(2):342--346, 1984.

\item{[8]}
R~M~Wald.
 Construction of solutions of gravitational, electromagnetic, or 
 other perturbation equations from solutions of decoupled equations.
 {\it Physical Review Letters}, 41(4):203--206, 1978.

\item{[9]}
S~A~Teukolsky.
 Perturbations of a rotating black hole. {I}. {F}undamental equations
  for the gravitational, electromagnetic, and neutrino-field perturbations.
 {\it The Astrophysical Journal}, 185:635--647, 1973.

\item{[10]}
G~F~Torres del Castillo.
 The {T}eukolski-{S}tarobinsky identities in type {D} vacuum
  backgrounds with cosmological constant.
 {\it Journal of Mathematical Physics}, 29(9):2078--2083, 1988.

\item{[11]}
G~F~Torres del Castillo.
 {K}illing spinors and massless spinor fields.
 {\it Proceedings of the Royal Society of London A}, 400:119--126,
  1985.

\item{[12]}
E~G Kalnins, R~G McLenaghan and G~C Williams.
 Symmetry operators for {M}axwell's equations on curved space-time.
 {\it Proceedings of the Royal Society of London A}, 439:103--113,
  1992.

\item{[13]} R Penrose and W Rindler. 
 {\it Spinors and Space-time}, volume 2.
 Cambridge University Press, Cambridge, 1986.

\item{[14]}
I~M Benn and R~W Tucker.
 {\it An Introduction to Spinors and Geometry with Applications in
  Physics}.
 IOP Publishing Ltd, Bristol, 1987.

\item{[15]}
J~A Thorpe.
 Curvature and {P}etrov canonical forms.
 {\it Journal of Mathematical Physics}, 10(1):1--7, 1969.

\item{[16]} 
B O'Neill.
{\it The Geometry of Kerr Black Holes.}
A K Peters, Wellesley, Massachusetts, 1995.

\item{[17]}
Kentaro Yano.
 Some remarks on tensor fields and curvature.
 {\it Annals of Mathematics}, 55(2):328--347, 1952.

\item{[18]}
M~Walker and R~Penrose.
 On quadratic first integrals of the geodesic equations for the type
  $\{22\}$ space-times.
 {\it Communications in Mathematical Physics}, 18:265--274, 1970.

\item{[19]}
N~Kamran and R~G McLenaghan.
 Symmetry operators for the neutrino and {D}irac fields on curved
  space-time.
 {\it Physical Review D}, 30(2):357--362, 1984.

\item{[20]}
I~M Benn.
 A unified description of null and non-null shear-free congruences.
 {\it Journal of Mathematical Physics}, 35(4):1796--1802, 1994.

\item{[21]}
P~Sommers.
 Properties of shear-free congruences of null geodesics.
 {\it Proceedings of the Royal Society of London A}, 349:309--318,
  1976.

\item{[22]}
J~N Goldberg and R~K Sachs.
 A theorem on {P}etrov types.
 {\it Acta Physica Polonica}, 12:13--23, 1962.

\item{[23]}
E Newman and R Penrose.
 An approach to gravitational radiation by the method of spin
  coefficients.
 {\it Journal of Mathematical Physics}, 3(3):566--578, 1962.

\item{[24]}
I~Robinson and A~Schild.
 Generalization of a theorem by {G}oldberg and {S}achs.
 {\it Journal of Mathematical Physics}, 4(4):484--489, 1963.

\item{[25]}
 O~Laporte and G~E Uhlenbeck, 
 {\it Physical Review}, 37:1380, 1931.

\item{[26]}
A~Nisbit.
 Hertzian electromagnetic potentials and associated gauge
  transformations.
 {\it Proceedings of the Royal Society of London A}, 250:250--263,
  1955.

\item{[27]}
E~Mustafa and J~M Cohen.
 {H}ertz and {D}ebye potentials and electromagnetic fields in general
  relativity.
 {\it Classical and Quantum Gravity}, 4:1623--1631, 1987.

\item{[28]}
B~Carter and R~G McLenaghan.
 Generalized total angular momentum operator for the {D}irac equation
  in curved space-time.
 {\it Physical Review D}, 19(4):1093--1097, 1979. 

\item{[29]}
L~P Hughston and L~J Mason.
 A generalised {K}err-{R}obinson theorem.
 {\it Classical and Quantum Gravity}, 5:275--285, 1988.

\item{[30]}
Shun ichi Tachibana and Toyoko Kashiwada.
 On the integrability of {K}illing-{Y}ano's equation.
 {\it Journal of the Mathematical Society of Japan}, 21(2):259--265,
  1969.

\item{[31]}
G~F~Torres del Castillo.
 Gravitational perturbations of algebraically special space-times via
 the {$\cal HH$} equation.
 {\it Journal of Mathematical Physics}, 27(6):1586--1591, 1986.

\item{[32]}
G~F~Torres del Castillo.
 Perturbations of solutions of the {E}instein-{M}axwell equations with a
 null background electromagnetic field.
 {\it Journal of Mathematical Physics}, 37(8):4053--4061, 1996.

\item{[33]}
G~F~Torres del Castillo.
 The separability of {M}axwell's equations in type-{D} backgrounds.
 {\it Journal of Mathematical Physics}, 29(4):971--977, 1988.

\item{[34]}
G~Silva-Ortigoza.
 Killing spinors and separability of {R}arita-{S}chwinger's equation
  in type $\{2,2\}$ backgrounds.
 {\it Journal of Mathematical Physics}, 36(12):6929--6936, 1995.

\item{[35]}
 E~G~Kalnins, G~C~Williams and W~Miller~Jr.
 Intrinsic characterization of the separation constant for spin one and
 gravitational perturbations in {K}err geometry.
 {\it Proceedings of the Royal Society of London A}, 452:997--1006, (1996).

\item{[36]}
S~V Dhurandhar, C~V Vishveshwara and J~M Cohen.
 Electromagnetic fields in space-times with local rotational symmetry.
 {\it Physical Review D}, 21(10):2794--2804, 1980.

\item{[37]}
S~V Dhurandhar, C~V Vishveshwara and J~M Cohen.
 Neutrinos in perfect fluid space-times with local rotational symmetry.
 {\it Physical Review D}, 26(10):2598--2610, 1982.

\item{[38]}
G~F~Torres del Castillo.
 Solution of the inhomogeneous equations of equilibrium for an isotropic
 elastic medium.
 {\it Revisto Mexicana de F\'{\i}sica}, 41(5):695--702, 1995.

\item{[39]}
I~M Benn and Jonathan Kress. 
 Force-free fields from Hertz potentials. 
 {\it Journal of Physics A:Mathematical and General}, 29:6295--6304,
 (1996) 

\end